\documentclass[conference]{IEEEtran}
\IEEEoverridecommandlockouts
\usepackage{cite}
\usepackage{amsmath,amssymb,amsfonts}
\usepackage{algorithmic}
\usepackage{graphicx}
\usepackage{nohyperref}
\usepackage{url}
\usepackage{textcomp}
\usepackage{float}
\usepackage{caption}

\usepackage{xcolor}
\def\BibTeX{{\rm B\kern-.05em{\sc i\kern-.025em b}\kern-.08em
    T\kern-.1667em\lower.7ex\hbox{E}\kern-.125emX}}
\usepackage{todonotes}

\usepackage{listings}

\lstdefinelanguage{Solidity}{
    keywords={pragma, solidity, contract, function, public, returns, uint, address, mapping, require, emit, event, view, pure},
    keywordstyle=\color{blue}\bfseries,
    ndkeywords={bool, string, uint256, int256, bytes},
    ndkeywordstyle=\color{teal}\bfseries,
    comment=[l]{//},
    morecomment=[s]{/*}{*/},
    commentstyle=\color{gray}\ttfamily,
    stringstyle=\color{red}\ttfamily,
    sensitive=true
}

\begin{document}

\title{A Practical Rollup Escape Hatch Design
}

\author{\IEEEauthorblockN{Francisco Gomes Figueira}
\IEEEauthorblockA{
\textit{Zircuit}\\
Madeira, Portugal \\
francisco@zircuit.com}
\and
\IEEEauthorblockN{Martin Derka}
\IEEEauthorblockA{
\textit{Zircuit}\\
Waterloo, ON, Canada \\
martin@zircuit.com}
\and
\IEEEauthorblockN{Ching Lun Chiu}
\IEEEauthorblockA{
\textit{Zircuit}\\
New York, USA\\
juno@zircuit.com}
\and
\IEEEauthorblockN{Jan Gorzny}
\IEEEauthorblockA{
\textit{Zircuit}\\
Toronto, ON, Canada \\
jan@zircuit.com}
}

\maketitle

\begin{abstract}
A rollup network is a type of popular ``Layer 2'' scaling solution for general purpose ``Layer 1'' blockchains like Ethereum.
Rollups networks separate execution of transactions from other aspects like consensus, processing transactions off of the Layer 1, and posting the data onto the underlying layer for security.
While rollups offer significant scalability advantages, they often rely on centralized operators for transaction ordering and inclusion, which also introduces potential risks. 
If the operator fails to build rollup blocks or propose new state roots to the underlying Layer 1, users may lose access to digital assets on the rollup.
An escape hatch allows users to bypass the failing operator and withdraw assets directly on the Layer 1.
We propose using a time-based trigger, Merkle proofs, and new resolver contracts to implement a practical escape hatch for these networks.
The use of novel resolver contracts allow user owned assets to be located in the Layer 2 state root, including those owned by smart contracts, in order to allow users to escape them.
This design ensures safe and verifiable escape of assets, including ETH, ERC-20 and ERC-721 tokens, and more, from the Layer 2. 
\end{abstract}


\section{Introduction}\label{sec:intro}

A Layer 2 (L2) blockchain network derives state from data on a different blockchain, called a Layer 1 (L1).
L2 networks have many nice properties that make them attractive~\cite{DBLP:conf/fc/GudgeonMRMG20}.
First, because L2 state is derived from and is posted onto an L1, L2 solutions inherit some security from their underlying L1 blockchain.
Second, they separate execution of state transitions (that is, processing of transactions) from consensus. 
In turn, they can process many more transactions per second than their underlying L1.
These advantages often go hand-in hand; for example, if an L2 posts its state root to L1, it can forgo the need for a consensus algorithm (as the consensus for valid state roots is whichever is most recent on the L1) and it can be free to post these roots as often as desired.
In order to derive a different L2 chain it is necessary to change the recorded L2 state on L1, a challenging and often prohibitively expensive task.
Coupled with well defined rules (implemented via smart contracts on the L1) as to when state roots are considered valid, L2 networks have a wide range of freedom for processing user transactions without giving up guarantees provided by the L1 networks they rely on.

Rollups are a popular L2 scaling solution.
A \emph{rollup} (also called a \emph{commit-chain} \cite{cryptoeprint:2018/642} or \emph{validating bridge} \cite{cryptoeprint:2021/1589}) is an L2 scaling solution where transactions are executed off-chain (i.e., off of the L1) and the batches of executed transactions, along with the resulting state roots of the L2, are posted on the underlying L1.
Batches of transactions are posted to the underlying L1, but not executed there.
To ensure that the transactions within a batch are executed correctly, the rollup also post state roots, each of which are the root of a Merkle tree \cite{DBLP:reference/db/Carminati18b} that contains all the state (i.e., variables) of the L2 blockchain.
Without loss of generality, we will use Ethereum as our L1 throughout this paper, for which there are more than 120 rollups (and similar systems) in development at the time of writing~\cite{l2_beat}.

In a rollup, transactions must be ordered for execution. 
Typically, a single actor, called the \emph{sequencer}, performs this task.
The sequencer orders rollup transactions and commits to the ordering by publishing the transactions. 
Later, the result of executing these transactions in the posted order is posted as a state root on the underlying L1.
Often, the actor performing these tasks -- sequencing and proposing -- is the same, especially as L2s are yet to include decentralized components.
This creates a single point of failure: if the operator disappears (or a bug results in the inability to publish a valid state root), L2 blocks and state updates will not appear on the underlying L1.
Critically, this means that transactions where users are withdrawing funds from these networks will also be missing, trapping user funds on the L2.

To address this issue and enable users to remove their assets from the rollup despite this point of failure, an ``escape hatch'' is necessary \cite{idealescapehatches}. 
This mechanism, unique to rollups, allows users to remove or ``escape'' their assets from the rollup and return them to their account on L1.
Without such a feature, digital assets (i.e., cryptocurrencies like Ether or ERC-20 tokens~\cite{erc20}) may be locked in the rollup system when state updates are not published on the L1: there will be no way to update the system to say that these assets are withdrawn.
Therefore, methods to unlock these funds and transfer important user state are necessary to overcome the risk of operator failure.
This is especially important as most rollups that are in use or under development are centralized, with a single operator.


An escape hatch provides partial \emph{liveness} for the rollup, that is, as long as the underlying L1 is operating, the funds on the L2 will always also be accessible.
This feature is therefore critical for users, but also non-trivial to implement, as these systems are complex.
Opting to fall back to centralized techniques to counter operational failure, escape hatches are not always present in rollups as teams focus on other key features.
While some rollups, such as Optimism~\cite{optimism}, have partially implemented this functionality in one way or another, many other rollups have not~\cite{l2_beat}.

Despite the importance of escape hatches, they have received limited attention in academic literature and real-world projects.
While escaping with Ether (ETH) and standardized transferrable assets such as ERC-20 and ERC-721 tokens~\cite{ERC-721} held in externally owned accounts (EOAs) is expected, it is unclear what escaping means for non-transferrable state recorded inside of smart contracts. 
Some smart contract applications, such as Uniswap~\cite{uniswapv2}, assume transferring assets to smart contracts, which makes escaping with such assets non-trivial. 
Existing literature offers no accepted standardization of existing escape hatch solutions, and provides no simplistic example of an architecture and implementation of an escape hatch in the centralized sequencer setting.

\emph{Our Contribution.} In this paper, we first provide a survey of existing approaches taken by rollups in development at the time of writing as well as related work.
Along the way, we establish key terms for defining this functionality in terms of practical end-user goals and consider challenges that may arise from adopting this design (Section~\ref{sec:background}).
Then we provide the first design of a practical escape hatch for a widely used rollup framework; namely, the OP Stack~\cite{opstack} (Section~\ref{sec:sysdesign}).
In doing so, we show that this functionality is feasible and useful with current rollup use cases (Section~\ref{sec:case-studies}).
Our design introduces the concept of a \emph{resolver}, which allows users to escape funds even if they are locked in L2 smart contracts.
We discuss the properties of our design (Section~\ref{sec:discussion}) before we conclude that this design is practical and accessible to rollups now (Section~\ref{sec:conclusion}).

\section{Background}\label{sec:background}

\subsection{Rollups}
A rollup can be broken down into several components\footnote{Other work like \cite{idealescapehatches, cryptoeprint:2021/1589} use different terms for these components, but each rollup has some component that performs these actions.}: a \emph{sequencer}, a \emph{state proposer}, and an (explicit or implicit) \emph{verifier}. 
A sequencer is responsible for ordering L2 transactions and committing, via a transaction to L1, to a batch of transactions to be executed. 
This batch is made up of L2 transactions.
A state proposer executes the transactions in a batch (in the order provided by the sequencer’s commitment) and computes new state roots which are written to L1. 
Verifiers in a rollup ensure that state roots are (eventually) correct. 
Depending on whether the rollup is \emph{optimistic} or \emph{zero-knowledge} (ZK), the responsibilities of these components may vary (see, e.g.~\cite{DBLP:journals/corr/abs-2404-16150}).

A sequencer orders transactions for the L2, building L2 blocks.
The source of these transactions may be a user of the rollup or the L1 smart contracts of the rollup.
As a result, sequencers are responsible for \emph{cross-chain} communication, and rollups implement a canonical blockchain \emph{bridge}.
A bridge is a system or protocol for taking assets or blockchain state from one blockchain to another.
As cryptographic assets cannot be literally moved from one blockchain to another, the bridge creates representations of assets on a \emph{source} blockchain on a \emph{destination} blockchain.
The bridge can be used to send Ether, messages or both; the rollup’s canonical bridge is necessary to bridge assets from the underlying L1 to the rollup itself.

L1 to L2 message passing is performed by interacting with the \texttt{L1Bridge} contract. 
These transactions are picked up by the sequencer and are forced to be included in the L2 on the protocol level. 
Sending ETH to this L1 contract it triggers the creation of the same amount of ETH on the L2. 

In a similar fashion, L2 to L1 message passing is performed by interacting with the \texttt{L2Bridge} contract. The hash of a message is saved on a storage slot in the bridge contract. 
Upon state root's publication and acceptance in the L1-residing \texttt{L2Oracle} contract, it becomes possible to submit and verify a Merkle proof that the withdrawal message was posted on the L2 by verifying the corresponding storage slot. 
ETH sent to this contract is burnt on L2 and unlocks the withdrawal of the corresponding amount of ETH that was previously deposited on the \texttt{L1Bridge} contract.

\subsection{Escape Hatches}\label{sec:subsec-escape-hatches}
This work differentiates between two different types of escape hatch functionality: \emph{forced inclusion} and \emph{forced withdrawal}.
This important distinction is necessary to make the notion of an escape hatch precise, and to differentiate the type of functionality necessary to overcome particular component failures.
In particular, forced inclusion mechanisms mitigate sequencer failure, while forced withdrawal mechanisms aim to mitigate failures with the rollup's state proposer. 

\emph{Forced inclusion} escape hatches serve to circumvent the sequencer. 
If the sequencer is offline or censors a transaction, the user can use such an escape hatch to include their transaction in a block.
This allows other honest actors running rollup nodes (often used to derive the state locally) to include the block and detect the censorship.

\emph{Forced withdrawal} escape hatches guarantee the ability to exit funds from the L2 without relying on the state proposer. 
Such functionality is relevant in case state proposers are offline, the state proposer's validity proofs have a bug, or the sequencer is censoring transactions.
In short, they aim to guarantee the successful withdrawal of assets from L2. 
Our design in Section~\ref{sec:sysdesign} is a forced withdrawal escape hatch.

A subtype of forced withdrawal escape hatches with the name \emph{idle proposer} is defined as follows. 
These escape hatches allow anyone to become a state proposer after the state has stopped progressing for a defined period of time, essentially making state proposers a permissionless role (so that anyone can fulfill this role). 
However, idle proposer escape hatches are only effective against state proposers that are offline - they might not be effective against malicious (colluding) state proposer(s). 
An important consideration is whether the protocol is ready to be permissionless when the idle proposer escape hatch is activated.
Note that the activation of this escape hatch may cause issues in the protocol, such as increasing the attack surface for so-called \emph{delay attacks}~\cite{delayattack}.

Any escape hatch is meant to be an emergency functionality. 
The usage of them should, in the best case, never occur and simply be a guarantee for users that they will be able to interact with the L2 and assets will not be locked on the L2. 
Nevertheless, if such an emergency state ever occurs, it is important that these escape hatches are accessible to users. 

\begin{table*}[bt]
\centering
\begin{tabular}{|c|c|c|c|c|c|}\hline
\textbf{Forced Inclusion} & \textbf{Forced Withdrawal} & \textbf{\shortstack{Permissionless \\ Escape \\ Sequencer Outage}} & \textbf{\shortstack{Permissionless \\ Escape \\ Proposer Outage}} & \textbf{Rollup Example}\\\hline
force via L1 & use escape hatch & yes & yes & zkSync Lite~\cite{zksync_lite_documentation}\\ \hline
self-sequence & self-propose & yes & yes & Optimism~\cite{optimism_sequencer}\\ \hline
enqueue via L1 & white-listed only with governance & no & yes with governance & zkSync Era~\cite{zksync_priority_queue} \\ \hline
No solution & white-listed only & no & no & Scroll~\cite{scroll} \\ \hline
\end{tabular}
\caption{Common escape hatch functionality as of Dec 2024 according to \href{www.l2beat.com}{L2Beat.com}.}
\label{tab:existed_solutions} 
\end{table*}

\subsection{Existing Solutions}\label{sec:existing-solutions}

As of December 2024, L2Beat~\cite{l2_beat} -- which tracks L2 development in the Ethereum ecosystem -- categorizes various forced inclusion and forced withdrawal escape hatch functionalities.
A summary of these approaches is presented in Table~\ref{tab:existed_solutions}.

L2Beat categorizes existing forced inclusion escape hatch mechanisms into four types: \emph{force via L1}, \emph{self-sequence}, \emph{enqueue via L1}, and \emph{no solution}.

The \emph{force via L1} approach (implemented in 10.42\% of rollups) enables users to submit \emph{withdrawal} transactions directly on L1  to ensure forced inclusion. 
Typically, this involves users providing a proof to the L1 contract, verifying their balance or state against the most recent state root. Upon successful verification and after a predetermined waiting period (often 7–14 days), users can withdraw their funds.

Under the \emph{self-sequence} approach (implemented in 72.91\% of rollups) as implemented by OP stack-based rollups, users can submit L2 transactions directly to the L1 contract when the sequencer becomes unavailable. The L1 contract validates these submissions and emits corresponding events, which other L2 nodes monitor and execute. 
This ensures that user-submitted transactions will eventually be included, even without a functioning sequencer.

The \emph{enqueue via L1} mechanism (implemented in 4.17\% of rollups) allows users to place their transactions into a priority queue on the L1 chain.
This guarantees that the sequencer cannot selectively skip these transactions, but their execution still depends on the sequencer’s availability.

The remaining (12.5\% of rollups) implement \emph{no solution}; their users have no recourse if the sequencer fails or censors transactions, leaving them unable to force inclusion.

L2Beat similarly classifies forced withdrawal escape hatch solutions into three categories: \emph{use escape hatch}, \emph{self-propose}, and \emph{white-listed proposer only} (denoted as \emph{cannot withdraw}).

The \emph{use escape hatch} (10.42\% of rollups) mechanism mirrors the force-via-L1 approach for inclusion. Here, users submit a valid proof, such as a Merkle proof or a zero-knowledge proof, to the L1 contract to demonstrate the validity of their balance or state. Systems that offer a \emph{force via L1} inclusion mechanism typically also have escape hatch for proposer outage. 
These approaches are not standardized and do not necessarily support escaping assets locked in L2 smart contracts.

The \emph{self-propose} mechanism (20.83\% of rollups) is the idle proposer mechanism of Section~\ref{sec:subsec-escape-hatches}, and permits anyone to become a proposer after a period of inactivity (commonly 5-7 days).
Under these circumstances, anyone can submit the updated state root to the L1 contract, ensuring progress in the L2 state. 
However, as there may be many interested parties in posting new state roots, there may be significant wasted effort in computing and proposing new state roots.

In systems restricted to \emph{white-listed proposer only} (68.75\% of rollups), funds remain locked if an authorized proposer does not step in.
Some solutions lie between these extremes, involving governance or shared authority. For example, in zkSync Era, only a whitelisted address can act as a proposer during a sequencer outage, but the governance system retains the power to modify or upgrade this arrangement.

\subsection{Related Work}\label{sec:related-work}
Koegl et al.~\cite{10.1145/3631310.3633493} have investigated potential attack vectors on rollups and proposed various countermeasures. 
They reiterate that to bolster rollup security, it is essential to establish a scalable, continuously accessible escape hatch, though they do not detail any designs.

Gorzny et al.~\cite{idealescapehatches} outline the ideal properties of an escape hatch: it should be modular—enabling flexible replacements—secure, and self-correcting, so that it does not rely on yet another escape hatch to mitigate vulnerabilities. Additional desirable features include arbitrary state escape, transaction efficiency, integrated functionality, broad applicability, and automated availability under certain conditions.
We comment on our design's properties in Section~\ref{sec:sysdesign}.

In \cite{DBLP:journals/corr/abs-2406-16219}, steps were taken to formally model possibly conflicting features of rollups, including escape hatch functionality.
They model the expected outcome of escape hatches interacting with other rollup functions but do not provide implementation details.

Our work fills in the gaps left by these papers by providing the details for a practical escape hatch mechanism that is broadly applicable. 
\section{Escape Hatch Design}\label{sec:sysdesign}
For simplicity (and without loss of generality), we describe our escape hatch design in the context of a ZK rollup that possesses a centralized sequencer that consumes and orders user transactions producing blocks that are added to the rollup blockchain.
These blocks are then posted on Ethereum Data Availability (DA) layer.
A ZK rollup proves the correct execution of the transactions posted and the validity of the state root through \emph{validity} proofs which are posted and verified directly on Ethereum. 
The state root of the most recently proved L2 block corresponds to the root a Merkle Patricia Trie (MPT) of account states, each of which includes Ether balance, nonce, code hash and storage root of another MPT whose leaves represent the storage slots of the account.
These assumptions simplify the exposition, e.g., by removing the need to consider long finality periods for state roots.
Nonetheless, our design is not limited to ZK rollups.

In normal operation, the rollup's canonical bridge architecture ensures that user funds cannot be forcibly removed back to L1.
In such cases, a withdrawal requires a state root posted to L1 with a valid withdrawal transaction that was recorded on the \texttt{L2Bridge} contract storage slot. 
The state root published on L1 can always be considered a sound inclusion proof as it is verified to be derived from a valid sequence of transactions by a validity proof.
We need to override this behavior when state roots fail to be posted, as their omission effectively locks digital assets within the \texttt{L2Bridge} contract.

\begin{figure*}[bt]
\captionsetup{justification=centering,margin=2cm}

    \centering
    \includegraphics[width=0.65\linewidth]{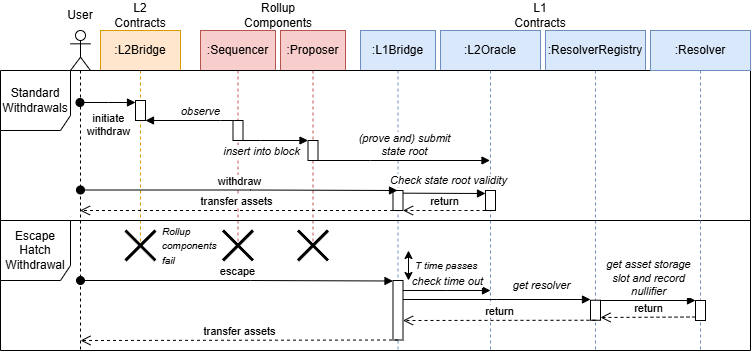}
    \caption{A comparison of a typical withdrawal process from an L2 (top) with our escape hatch process (bottom).}
    \label{fig:escape-comp}
\end{figure*}

We define an automatic trigger that allows users to escape with their funds, bypassing the normal message passing flow of the bridge contracts if state roots fail to be posted (or be successfully verified).
To do this, we define a time limit $T$, after which if no new state root has been posted, users are permitted to use the escape hatch and bypass the bridge.

Once time $T$ has passed and escaping is enabled, we need to determine a way to retrieve the most recent valid state of the rollup. 
All blocks produced are posted on Ethereum DA; however, this data is not verified and may contain invalid blocks (e.g., the sequencer may be online but inserting arbitrary minting transactions), making it an unreliable source for the most recent state. 
Thus, we \emph{must} rely on the most recent state root posted on Ethereum which was accompanied by a validity proof. 
From this root, we can generate Merkle proofs of the entire state of the rollup, including users' ETH balances and storage states for all smart contracts at the time this state root was constructed.
This setup ensures that we can prove the account state based on the most recently valid state. 

However, there is a challenge in figuring out where the asset is within the state root.
The escape hatch must be able to prove that the asset is included in the MPT and that the caller is in fact entitled to that digital asset.
To do this, we introduce the concept of a \emph{resolver} contract.
A resolver contract takes a state root, an asset contained on L2, a target L2 address representing an L2 smart contract, and returns a storage slot's location for the supplied address within the storage slot of the supplied L2 contract within the L2 for the caller of the escape hatch.
Discussion on how to implement and choose a resolver contract can be found in Section~\ref{registering resolvers}.
In short, it is responsible for finding the location of the asset on the L2 that the user wishes to escape.

Finally, we must ensure that users can only escape their assets once. To do this, we create a nullifier, a value that can be used to prevent double spending, to track whether an account has escaped with its assets already; this prevents bad actors from removing funds multiple times.
To ease the implementation of the nullifier, we do not allow users to partially escape any particular asset: if a user is to escape an asset at a particular leaf of the state root, they must take their entire share of that value (e.g., they cannot take only 50\% of the ETH they have locked on the L2).

The escape hatch process is compared to a typical L2 withdrawal in Figure~\ref{fig:escape-comp}.
In the following sections, we outline in detail the steps necessary to escape various assets by suggesting the functionality of common resolver contracts.

It is important to notice that in this design once a single escape is performed, the rollup would require a hard fork to be restarted since the amount of assets present on the bridge in L1 would be different from the amount assets recorded in the L2 state.
However, this is likely undesirable: as assets may have escaped the L2 from smart contracts on it, those smart contracts may have had internal variables tracking those assets which would be inconsistent on the new blockchain fork.

\begin{lstlisting}[language=Solidity, caption={A call to get a users account balance from an L2 node.}, label={lst:javascript1},float]
const proof = await rpc.send('eth_getProof', [userAddress, [], targetBlockNumber]);
\end{lstlisting}

\subsection{Ether (ETH)}\label{sec:ether-escape-intro}
ETH is the native asset on the majority of rollups, meaning its balance is recorded in the account state’s MPT leaf node. 
Since ETH is not dependent on the storage state of a smart contract, it is the simplest asset to escape, and does not require a resolver contract. 
The user simply needs to provide a Merkle proof of their account state based on the most recently posted state root.
The user would first request a Merkle proof of their account state from an L2 node as in Listing~\ref{lst:javascript1} (which can rederive the L2 state from the DA and posted state roots and generate such a proof).
This Merkle proof is then used as input to a function call to the \texttt{L1Bridge} contract, the bridge contract calls the \texttt{L2Oracle} contract where the L2 state roots are recorded, and obtains the most recent valid state root together with the timestamp at which it was posted. The bridge contract verifies the validity of the provided Merkle proof using the state root, and ensures that the timestamp of the last output root is old enough to enable triggering an escape.
If all checks are successful, the ETH is transferred back to the user, and a nullifier is recorded on the bridge contract that will disallow another escape by the same user. 

\subsection{Other Assets} \label{other assets}
Other assets which are defined using the ERC-20, ERC-721, or ERC-1155~\cite{ERC-1155} standards have the user balance saved in a smart contract storage slot. 
This is an issue for escaping the assets as we need to define a way to locate the correct storage slot where the user balance (or \texttt{tokenId}) is located. 
Even within the same standard there are multiple implementations; e.g., for ERC-20 there are two commonly used implementations (one by by Openzeppelin~\cite{openzeppelin} and another by transmissions11~\cite{transmissions11}) which are both compliant with the ERC-20 but store user balances in different storage slots.

In addition, users can keep their assets not in Externally Owned Accounts (EOA) but in smart contracts such as Automated Market Makers (AMMs), lending and borrowing markets, among others, which enable users to earn passive income from their assets (see, e.g.,~\cite{DBLP:journals/corr/abs-2404-11281} for an introduction to Decentralized Finance, a.k.a.~DeFi, protocols).
The use of so-called smart wallets which provide users access to new functionalities such as wallet recovery and session keys for a better user experience is also becoming increasingly popular.
Each implementation of these smart contracts possess different logic and therefore different ways are necessary to determine how much a user on L1 would be entitled to for any given asset in case of an escape.

In these cases, we can make use of resolver contracts.
These contracts implement the logic required to determine how many (or which) assets a given user is entitled to within an L2 smart contract. 
The resolver contract should assume it can read the value of any arbitrary storage slot and through that be able to return the assets and amount that a user can escape with.
In order to be able to use the resolver contract, they must be deployed onto Ethereum and recorded in a registry that the rollup \texttt{L1Bridge} can consult in order to allow the users to remove their assets. 
Ideally the resolver contract should be registered by the L2 contract that holds the assets, which would make the \texttt{RegistryResolver} contract can be as simple as in Listing~\ref{lst:solidity_exampleA}; see also Section~\ref{registering resolvers}.
The \texttt{RegistryResolver} contract mains a list of address for specific resolver smart contract, each of which implements logic to unlock assets in a given L2 smart contract.

\begin{lstlisting}[language=Solidity, caption={An example \texttt{ResolverRegistry} contract, used to record resolvers used in the escape hatch.}, label={lst:solidity_exampleA},float]
contract ResolverRegistry{
    mapping(address => address) public resolvers;

    CrossDomainMessenger immutable messenger;

    constructor(address _messenger) {
        messenger = CrossDomainMessenger(_messenger);
    }
/// @dev This function can only be used by performing a cross chain call from L2 cross domain messenger.
    function setResolver(address _resolver) external {
        require(msg.sender == address(messenger));
        address sender = messenger.xDomainMessageSender();
        resolvers[sender] = _resolver;
    }
}

\end{lstlisting}

\subsection{Registering Resolvers}\label{registering resolvers}

It is important to determine who can register a resolver for an escape situation.
Malicious actors may wish to register resolvers which do not assist honest users to escape their L2 assets -- the resolver contract may implement arbitrary logic, and it therefore may be possible to implement resolvers which steal user funds from the L2 upon an escape attempt.
We propose two ways to register resolvers: while the L2 smart contract is accessible on the L2, and after operator failure.

\subsubsection{Live Contract Resolver Registration}\label{subsubsec-live-contract-registration}
One way to ensure that a resolver is registered for an L2 smart contract is to have the L2 smart contract register its own resolver.
In particular, L2 smart contracts can implement a function to send a message to e.g., the \texttt{L2Bridge} contract to register an L1 address which is to be used as the resolver in case the L2 fails to operate.
This requires the L1 address of the resolver address to be known, which requires the resolver to be deployed on L1.

Since resolvers can implement arbitrary logic, resolvers may introduce security risks.
As with any smart contract there is a possibility of existing an exploit in the contract logic that may affect the users. 
There is also a chance that the logic implemented is malicious towards users, however it must be taken into account that the deployers of the resolver contract will in principle be the same deployers of the L2 contract.
Therefore the trust assumption of the resolver is inherited from the L2 contract in this approach. 
In turn, this lets users of the L2 smart contract inspect the resolver prior to using the application.
This allows them to determine if the resolver contract itself poses any risk to their funds in the event the L2 fails to operate as planned.

\subsubsection{Post-Operator Resolver Registration}\label{subsubsec-post-failure-registration}
Operator failure is an unlikely event to occur as the sequencer runner is incentivized to keep the network alive in order to earn the transaction fees earned on the rollup. By the same logic, smart contract developers on L2 are not incentivized to register resolver contracts prior to operator failure as it consumes developing time to create a feature that is unlikely to be used. Our design therefore allows for  resolver registration in a period after operator failure, to allow L2 smart contract developers to defer this development cost to a time when it becomes necessary.

In this case the most trusted entity would be the deployer of the contract on L2 as they should be the developer behind the protocol.
A contract's address is determined by the \texttt{create} opcode, which is computed as $H(rlp([a,n]))$ where $H$ is the Keccak256 hash function, $rlp$ is the Recursive-Length Prefix (RLP) serialization used by Ethereum \cite{rlp}, $a$ is the address of the deployer, and $n$ is the nonce of the transaction used to deploy the smart contract.
If the \texttt{create2} opcode is used instead then the address is computed as $H(0\text{x}FF||a||salt||H(bytecode))$ where $H$ is the Keccak256 hash function, $||$ is the string concatenation function, $a$ is the address of the deployer, $salt$ is a random value chosen by the deployer, and $bytecode$ is the bytecode of the deployed contract.
To allow registering resolvers after the fact we can use the logic shown in Listing~\ref{lst:deployer_address}.

\begin{lstlisting}[language=Solidity, caption={Sample (pseudo-code) implementations guarding the registration of resolvers, necessary to avoid malicious actors from adding resolvers in the hopes of eventually stealing funds.}, label={lst:deployer_address},float]
function canRegister(uint256 nonce, address l2ContractAddress) public view returns (bool) {
  bytes memory data = RLP(msg.sender, nonce);
  return keccak256(data) == l2ContractAddress;
}
function canRegister2(uint256 nonce,bytes32 salt, bytes32 bytecodeHash address l2ContractAddress) public view returns (bool) {
  bytes memory data = abi.encodePacked(0xFF,msg.sender, salt, bytecodeHash);
  return keccak256(data) == l2ContractAddress;
}
\end{lstlisting}

However this approach does not cover the case where the contract was deployed using a factory contract. For such scenarios we propose that a standard should be used to include the deployer address in the salt of \texttt{create2} when deploying a contract using a factory contract. This way it is simple to verify the deployer of the contract and grant them permissions to register a resolver (later).
Specifics of this case are left as future work.

This case may introduce arbitrary risk to users who wish to escape funds from a protocol as the resolver logic is not known to users when they interact with the corresponding L2 smart contract.
Therefore we propose the implementation of another time lock: resolvers which were registered via an L2 message may be activated after time $T$, but those registered post-operator failure may need to wait until $2T$ time has passed.
While this does not mitigate malicious resolver code, it increases time to investigate and prioritizes the more trustworthy resolvers registered prior to the operator failure.
\section{Case Studies}\label{sec:case-studies}

In this section we provide example resolvers for well-known Ethereum tokens (ERC-20 in Section~\ref{sec:case-study-20} and ERC-721 in Section~\ref{sec:case-study-721}) and outline the steps required to support wallet contracts (Section~\ref{sec:case-study-wallet}) and escape funds in the more complicated example of Uniswap v2 (Section~\ref{sec:case-study-uniswapv2}).

\subsection{ERC-20 Tokens}\label{sec:case-study-20}

Unlike ETH, ERC-20 balances are stored inside a storage slot of a smart contract, and as described in Section~\ref{other assets} the slot for where the user balance is stored depends on how the standard is implemented. 
Therefore in order to allow users to escape ERC-20 there is a need for a resolver contract that will return the storage slot where the user balance is located.
Therefore in an ideal scenario ERC-20 tokens that allow for escape would implement a function that will enable them to register a resolver on L1 by means of sending a cross chain transaction.
The resolver for an ERC-20 contract can be implemented as in Listing \ref{lst:solidity_example2}.

\begin{lstlisting}[language=Solidity, caption={A sample resolver  implementation for ERC-20 tokens.}, label={lst:solidity_example2},float]
  function getERC20Slot(address _user) external pure returns (bytes32) {
 return keccak256(abi.encode(uint256(uint160(_user)), uint256(BALANCES_SLOT)));
 }
\end{lstlisting}

This enables a way to verify in which storage slot we can find a given user balance. It is also the case that any ERC-20 that utilizes the same slot for the balances can reuse the same resolver contract and register it on the registry.
As it is not realistic for every ERC-20 to register a resolver we can also enable default resolvers (parameterized for each ERC-20 token address on the L2) for the most common balance slots for ERC-20 implementations on the L2.

The process of escaping ERC-20 starts with the user requesting a Merkle proof of the ERC-20 account state, and a Merkle proof of the slot corresponding to the user token balance to a L2 node.
These two Merkle proofs are then used as input to the escape function in the \texttt{L1Bridge} contract. 
The bridge contract will perform a call to the \texttt{L2Oracle} contract to obtain the most recent L2 state root together with the timestamp at which it was posted. 
It would then check the timestamp to be old enough to enable escape, and verify the Merkle proof of the ERC-20 account state using the state root. By verifying the account state it now has access to a sound storage root enabling the verification of any storage slot of the contract. 
In order to determine which storage slot the user balance is located it needs to call a resolver as introduced in Listing \ref{lst:solidity_example2}, a call will be performed to the \texttt{ResolverRegistry} contract to check if the ERC-20 has registered a resolver.
If the resolver was registered, then it would be called to obtain the user balance storage slot location; otherwise it would call the default ERC-20 resolver and attempt to obtain the user balance storage slot location. Then it would verify the provided storage Merkle proof for the balance storage slot location obtained from the resolver, using the storage root present in the ERC-20 account state.
If all checks are successful, the ERC-20 would be transferred to the user, and a nullifier would be set disallowing the user to escape the same asset again.

\subsection{ERC-721 Tokens}\label{sec:case-study-721}

Escaping ERC-721 tokens is very similar to escaping ERC-20, as the ERC-721 token contract would need to deploy and register a resolver contract on L1. However, instead of providing the slot with the balance of a given user, the resolver contract would instead provide the storage slot where the owner of a given \texttt{tokenId} is registered.
The resolver of an ERC-721 can be implemented as in Listing~\ref{lst:solidity_example3}.

\begin{lstlisting}[language=Solidity, caption={A sample resolver  implementation for ERC-721 tokens.}, label={lst:solidity_example3},float]
  function getERC721Slot(uint256 _tokenId) external pure returns (bytes32) {
        return keccak256(abi.encode(_tokenId, uint256(OWNERS_SLOT)));
    }
\end{lstlisting}

This enables us to find the slot where the owner of a given \texttt{tokenId} is stored. As in the ERC-20 any number of ERC-721 can reuse the same resolver contract as long as the owner's slot is in the same position.
As for ERC-20 tokens, default resolvers can be deployed by analyzing the most commonly used owner slots for ERC-721 in the rollup which enables escaping without any action required from the token deployers.

The escape process is then similar to the one described in Section~\ref{sec:case-study-20}: simply modify the resolver used to get the relevant storage slot.

\subsection{Escaping with Wallet Contracts}\label{sec:case-study-wallet}

Wallet contracts are becoming more common as they provide greater safety (i.e., multi-signature wallets to prevent single points of failure) and to improve user experience. However, the address of these smart wallets is often not preserved across chains. 
The same user could deploy a wallet contract on both L1 and L2 and not necessarily have the same address on both, or use a wallet contract on L2 but an EOA on L1. Therefore, when attempting to escape assets, it would not be possible to determine that the assets of the user on L2 belong to the address on L1.

Given the increasing popularity of smart wallets, it makes sense for the escape hatch to have a dedicated solution for this scenario. We propose enabling the registration of delegate accounts, similar to how a deployer of a contract on L2 can register a resolver; the deployer of a smart wallet can also register a delegate that is enabled to escape assets in place of the smart wallet address in L2.

With this solution, users can now escape all their assets in the same way they would escape with an EOA, they just perform an additional transaction to register the address that can be used to escape in place of the L2 smart wallet address (see Section~\ref{subsubsec-live-contract-registration}).

\subsection{Escaping from Uniswap V2}\label{sec:case-study-uniswapv2}

Uniswap v2 \cite{uniswapv2} is an AMM that holds two tokens (token $X$ and token $Y$) in the $Pool$ smart contract. It allows anyone to withdraw token $X$ from the pool, by depositing an amount of token $Y$ determined by the ratio of total amount of token $X$ and token $Y$ in the pool. 
Assets are provided to the pool by liquidity providers, who receive so-called \emph{LP tokens} to represent their share of the pool. The amount of $X$ and $Y$ tokens that can be withdrawn by burning LP tokens depends on the ratio $X$ to $Y$ tokens in the pool, and the total amount of existing LP tokens.

In the case of prolonged operator failure the tokens deposited into the $Pool$ contract by the liquidity providers would be trapped there.
Since there is no single owner of tokens we need a way to determine how many $X$ and $Y$ tokens each liquidity provider is entitled to.
For this scenario and other DeFi protocols we need resolver contracts to be able to implement arbitrary logic in order to be able to correctly distribute the assets placed in the contract to its users.
Since it is also possible to access and verify arbitrary storage slots one can design a resolver contract that is able to correctly determine the amount of tokens each liquidity provider can remove from a Uniswap v2 $Pool$ contract.

We propose that a resolver for Uniswap v2 implement the following steps:

\begin{enumerate}
\item Verify the addresses of tokens $X$ and $Y$, that the pools used by reading the storage slots where they are stored.
\item Get the balance of the $X$ and $Y$ tokens held by the pool.
This can be done by applying the methods described on section \ref{sec:case-study-20} assuming that the ERC-20 tokens in the pool have a resolver deployed, or are compatible with a default ERC-20 resolver.
\item Check the total amount of LP tokens that were minted by the pool.
This can be done by reading the $totalSupply$ variable's storage slot.
\item Verify the amount of LP tokens that the user possessed. 
Once again, this information is stored in a storage slot in the pool contract.
\item Using the values obtained from the storage slots we can use the logic present on the $burn$ function of the $Pool$ contract to determine the amount of $X$ and $Y$ tokens any given liquidity provider is entitled to. 
\item Transfer the assets from the bridge to the user and save a nullifier so that the same user cannot remove the assets a second time.
\end{enumerate}

A resolver that implements these steps allows liquidity providers of Uniswap v2 to escape their assets back to L1. However, this capability is not limited to Uniswap v2, as resolvers can verify arbitrary storage slots and implement arbitrary logic it is also possible to develop resolvers for arbitrary smart contracts. 

\section{Discussion}\label{sec:discussion}

Recall that \cite{idealescapehatches} highlights some ideal  properties of escape hatches. 
We start by commenting on the properties we have in this design.
First, our design is modular, secure, and correcting.
In particular, resolvers can have custom logic tailored to escape needs and may even be placed behind so-called \emph{proxy} patterns~\cite{DBLP:journals/ese/EbrahimiAOH24}.
Resolvers for well-known digital assets are simple and do not increase the attack surface of the rollup if they are gated behind a sufficiently long time out and do not use a proxy pattern.
By design, they also allow the locked funds to be released on the \texttt{L1Bridge}, correcting the issue of trapped funds.
As resolvers can implement arbitrary logic, this design allows arbitrary digital assets to be escaped (or simply to be proven as existing on a slot in an offline L2).
Our design is not entirely built-in nor global as L2 smart contracts do require custom functionality, but it allows for common tokens to be escaped without additional effort.
It is also automatically available after a given timeout, provided the necessary resolvers are implemented.
Finally, while it requires users to send transactions, advanced implementations could allow for multiple assets to be simultaneously escaped.
Furthermore, since it does not attempt to restart the chain, it will not have wasted effort like that of idle-proposed approaches, which are also not transaction efficient.
In short, our design is a compromise of ideals in order to provide practical security without significant development barriers for rollup developers.

The design is also based on using a MPT to produce state roots.
If another rollup uses another hash function in their MPT (e.g., a more ZK-friendly function like Poseidon), our approach still works. However, it may become infeasible to use in practice since Ethereum may lack these ZK friendly hash functions as precompiles, which may  increase the cost of an escape on such a chain.
One way to mitigate this is to perform resolver computations off-chain in a ZK proof system, and simply require the resolver to verify the corresponding proof.
This may allow gas savings in some situations like the one above.
However, this approach can be generalized: in the event that an escape requires several complex steps (e.g., more than as in Section~\ref{sec:case-study-uniswapv2}), the resolver logic can be placed in a ZK proof system.
This further increases the modularity of our design.

This modularity also means that L2-native assets may be able to be entirely recreated on L1, as the resolver may mint new tokens on the L1. 
This should be done with care to avoid creating multiple copies of the same token, but may be desirable in some situations.
For example, a case that requires special attention is Wrapped ETH (an ERC-20 representation of ETH).
This token is usually created by bridging ETH and then wrapping it on L2; therefore, when escaping WETH this token may not be present on the bridge in the same amount as it exists on L2, requiring users to receive ETH instead.

Finally, the choice of timeout $T$ should be carefully considered by any rollups using this approach.
A time shorter than any upgrade delay restrictions on rollup contracts may mean that users are forced to use the new rollup contracts (which may introduce risk), while a time longer than such a delay may effectively prevent the rollup operators from simply upgrading a contract to correct any issues that caused the chain to stop posting state roots in the first place.
Each rollup should consider its community's needs when choosing these parameters.

\section{Conclusion and Future Work}\label{sec:conclusion}

In this work we have described a practical design that allows users to escape a rollup in case of operator failure. 
As escaping this design would break the state of the chain it should only be used in emergency situations when new state roots are not being posted.
As centralized rollups are developing there are concerns about censoring users and operator failure, rendering users unable to withdraw rollup funds.
We presented a design which is effective and easy to implement on existing rollups.

In the future, escape hatches based not on providing Merkle proofs but by using a general purpose ZK Virtual Machine (ZKVM) could result in a simpler design.
For example, smart contracts could forgo the need to deploy an accompanying resolver contract and instead implement a function on the contract itself on L2 that would contain the logic needed to determine how many assets a user is entitled to.
By using a ZKVM and the most recent state root, it is possible to create a proof of the return result of that function which corresponds to the amount of assets a user can escape with.
However, this may not be necessarily more advantageous than our approach and will likely also require one escape transaction per digital asset and user pair.
Moreover, it may require expensive hardware to execute and additional modifications to canonical bridge logic, unlike the approach we have outlined.
This is an interesting direction for future work.

\bibliographystyle{ieeetran}
\bibliography{ref.bib}

\end{document}